\documentclass[preprintnumbers, prd, onecolumn, floatfix,
preprintnumbers,showpacs, letterpaper, superscriptaddress,nofootinbib]{revtex4}
\usepackage{graphicx,epsfig}
\usepackage{amsmath}
\usepackage {amssymb}
\usepackage{subfigure}
\usepackage{color}

\begin{document}

\title{Some Astrophysical Aspects of a Schwarzschild Geometry Equipped with a Minimal Measurable Length}

\author {\textbf{Mohsen Khodadi}}
\email{m.khodadi@stu.umz.ac.ir}
\affiliation{Department of Physics, Faculty of Basic Sciences,\\
University of Mazandaran, P. O. Box 47416-95447, Babolsar, Iran}

\author {\textbf{Kourosh Nozari}}
\email{knozari@umz.ac.ir (Corresponding  Author)}
\affiliation{Department of Physics, Faculty of Basic Sciences,\\
University of Mazandaran, P. O. Box 47416-95447, Babolsar, Iran}

\author{\textbf{Anahita Hajizadeh}}
\email{an.hajizadeh@umz.ac.ir}
\affiliation{Department of Physics, Faculty of Basic Sciences,\\
University of Mazandaran, P. O. Box 47416-95447, Babolsar, Iran}

\begin{abstract}
By considering a deformation of the Schwarzschild metric in the presence of a minimal measurable length which still respects the equivalence principle, we study
corrections to the standard general relativistic predictions for some astrophysical phenomena such as stability of circular orbits of
black hole accretion disks, redshift of black hole accretion disks, gravitational tidal forces and the geodetic drift rate.
We use the \emph{Gravity Probe B} data to see robustness of our results. Our analysis shows also that the
relevant deformation parameter $\varepsilon$ which has a geometric origin, plays the same role as the charge to mass ratio, $\frac{e}{m}$
in the Reissner-Nordstr\"{o}m metric.
\end{abstract}
\pacs {04.60.-m, 04.60.Bc, 04.70.Dy}

\maketitle

\section{Introduction}

Despite the fact that general relativity (GR) is a well understood theoretical framework
with some strong empirical supports, it breaks down at very small length scales and a quantum theory of
gravity is required. In fact, to achieve a coherent picture of the Universe
from its beginning until today, we need a unified framework of general relativity and quantum filed theory.
Current speculative approaches to quantum gravity such as string theory, loop
quantum gravity, deformed spacial relativity and also quantum physics of black holes, phenomenologically
steer one toward replacing the concept of point in spacetime geometry with an \emph{``invariant minimal
length''}. This feature can be encoded at the high energy regime by generalizing the standard
Heisenberg uncertainty principle to the so called \emph{``gravitational (generalized) uncertainty principle'' (GUP)} \cite{Hossenfelder:2013}.

Even though quantum gravity is still under development, all proposed approaches so far such as loop quantum gravity
\cite{Rovelli:1995, Ashtekar:1997}, string theory \cite{Gross:1988, Amati:1989}, doubly
special relativity \cite{Amelino:2002,Magueijo:2002} and noncommutative geometry \cite{Seiberg:1999,Connes:2000}
require existence of an invariant minimum length on the order of the Planck length, $\ell_{pl}\approx10^{-35}m$.
Astonishingly, in the light of taking such a fundamental length in physics of black holes into account, some
issues such as the \emph{information loss paradox}, have been alleviated in recent years \cite{Maggiore:1993}.
From an experimental standpoint, the fundamental length may be accessible by test particles of short wavelength (high energy)\cite{Pikovski:2012, Nozari:2011}.
However, in this regime, the spacetime structure significantly is affected by the quantum gravitational effects
of the test particles. Technically speaking, beyond the Planck scale due to quantum fluctuations of the background metric, topology of spacetime
changes in essence and spacetime manifold turns to be discrete. If one incorporate the gravitational effects in quantum phenomena (remember the Heisenberg electron microscope thought experiment), then the standard uncertainty
principle changes to the generalized uncertainty principle (GUP)\cite{Kempf:1995, Kempf:1997} (see also \cite{Nozari:2012}). From an algebraic
viewpoint, the generalization of the standard Heisenberg commutators between position and momentum in Hilbert space guarantees appearance of
an invariant minimum length. We note that in the framework of polymer quantization a similar commutator relations can be obtained as well (see for instance  \cite{Hossain:2010}-\cite{Nozari:2015} and references therein). Focusing on existing literature, we see that in the context of GUP admitting just a minimum
measurable length one deals with almost two types of representations. Albeit this distinction has in fact a phenomenological origin
towards measuring the size of modifications appeared in different kinds of GUP by applying explicit
bounds on the relevant GUP deformation parameter.
As the first type, one can follow
works such as \cite{Baru:1999}-\cite{Ali:2011} where a particular representation of the operators in the modified
Heisenberg commutator, $[\hat{X},\hat{P}]=i\Big(1+\beta\frac{\hat{P}^2}{m_{pl}^2}\Big)$ has been considered.
We note that the phenomenological outcomes risen from this version (specifically, upper bounds
extracted for the modified dimensionless parameter $\beta$ which is expected to be of the order of unity)
are not so interesting theoretically since they are \emph{representation dependent}. In other words,
within the mentioned approach, there are many possible representations for operators $\hat{X}$ and $\hat{P}$. The second type
has been suggested through the deformation of the classical Newtonian mechanics. More technically, it has been
introduced via modifying the standard Poisson brackets as $\{x,p\}=1+\beta_0p^2$ so that in a sense it looks like the quantum commutator, $[\hat{x},\hat{p}]=
i(1+\beta_0\hat{p}^2)$ where $\beta_0=\frac{\beta}{m^{2}_{Pl}}$ (see for instance \cite{Cheng:2002}-\cite{Pedram:2012}).
A serious criticism about this version of GUP is that while GR is deformed at the order of $\beta$ with this GUP, the Newtonian
Poisson brackets remains unchanged. To be more clarified, in Appendixes A and B of Ref. \cite{Fabio:2015}, it has been shown in details that ``Equivalence Principle'' can be violated in the context of \emph{``$\beta$-deformed Newtonian mechanics Poisson brackets''}, while this does not occurs within the
\emph{``Newtonian limit of $\beta$-deformed GR''}. In other words, this version of the deformed Heisenberg algebra in the limit $\beta\rightarrow0$
reproduces the Newtonian
classical mechanics and not GR. Therefore some commutators containing corrections relevant to GR also must be proposed additionally.
In Ref. \cite{Fabio:2015} a new version of GUP with minimal length has been realized in which rather than focusing on given
representations of canonical operators $\hat{X}$ and $\hat{P}$ or modifications of the classical
Newtonian mechanics, one starts from a pure quantum effect known as the \emph{``Hawking evaporation"}.
Interestingly, in the mentioned reference \cite{Fabio:2015}, the authors were able to explicitly link the minimal length
modification of the Schwarzschild geometry to the uncertainty relation independent to any given representation of commutators.
Also in this new version, without recourse to the Poisson brackets and Newtonian mechanics,
the standard GR automatically retrieves in the limit $\beta\rightarrow0$. A prominent feature of the new approach is that the
equation of motion of test particle yet obeys the standard geodesics equation. \\

With these preliminaries, in this paper we study some phenomenological aspects arising from incorporation of a minimal measurable
length into Schwarzschild geometry according to the metric introduced in \cite{Fabio:2015}.
The interesting feature of this metric is that it is equipped with a minimal measurable length whilst it still respects the equivalence principle.
For this purpose, we have devoted sections II to V to four noteworthy astrophysical phenomenons respectively as: 1) The stability status
of circular orbits of a black hole accretion disk, 2) The redshift of the black hole accretion
disk, 3) Gravitational tidal forces and 4) The geodetic drift rate. Then the paper follows by summary and conclusion. We use the
signature $(+,-,-,-)$ through this paper.

\section{Stability of Circular Orbits of Black Hole Accretion Disk}
In this section we derive an energy condition (and subsequently a shape equation) in Schwarzschild
geometry equipped with a minimal measurable length \cite{Fabio:2015}. Then we study the effect of
minimal measurable length on the stability of circular trajectories of massive gas particles in black hole accretion disk.
For this purpose, we firstly extract the effect of the minimal length
correction terms for the geodesics in the GUP-deformed Schwarzschild geometry.
By avoiding details (as have been reported in \cite{Fabio:2015}), a deformed Schwarzschild
metric with a minimal measurable length is as follows

\begin{equation}\label{e2-1}
dS^{2}=c^2F(r)dt^2-F^{-1}(r)dr^{2}-r^2(d\theta^2+
\sin^{2}\theta d\phi^2),~~~F(r)\equiv \Big(1-
\frac{2\mu}{r}+\varepsilon\frac{\mu^2}{r^2}\Big),~~~
\mu\equiv\frac{GM}{c^2}\,.
\end{equation}
With this deformed line element, the GUP-deformed Hawking temperature can be recovered interestingly.
The third term in $F(r)$ can be considered as a perturbation term due to incorporation of a natural cutoff
as a minimal measurable length. Note that $\varepsilon$ is of the order of unity, that is, $|\varepsilon|\leq1$ with possibly
negative sign. In this paper we consider only the leading order modification in
the dimensionless parameter $\varepsilon$. Even for the case that $\varepsilon$ is very close to unity, since $\frac{GM}{r}$ is
usually small (for instance, it is $\approx10^{-5}$ at the surface of Sun), our approximation is justified.

The deformed Lagrangian relevant to line element (\ref{e2-1}) is given by
\begin{equation}\label{e2-2}
L=c^2F(r)\dot{t}^2-F^{-1}(r)\dot{r}^{2}-r^2
(\dot{\theta}^2+\sin^{2}\theta \dot{\phi}^2),
\end{equation}
where $L=g_{\mu\nu}\dot{x}^{\mu}\dot{x}^{\nu}$
with $\dot{x}^{\mu}\equiv\frac{dx^{\mu}}{d\sigma}$ and $\sigma$ is a parameter along the geodesics.
After substituting the above modified Lagrangian
into the Euler-Lagrange equations,
\begin{equation}\label{e2-3}
\frac{d}{d\sigma}\bigg(\frac{\partial L}{\partial\dot{x}
^{\mu}}\bigg)-\frac{\partial L}{\partial x^{\mu}}=0~,
\end{equation}
we arrive at the following geodesic equations for Schwarzschild spacetime
deformed by a minimal length
\begin{eqnarray}\label{e2-4}
\begin{array}{ll}
\Big(1-\frac{2\mu}{r}+\varepsilon\frac{\mu^2}{r^2}\Big)\dot{t}=k~,\\\\
\Big(1-\frac{2\mu}{r}+\varepsilon\frac{\mu^2}{r^2}\Big)^{-1}\ddot{r}
+\Big(1-\frac{2\mu}{r}+\varepsilon\frac{\mu^2}{r^2}\Big)^2\Big(\frac
{\mu}{r^2}-\varepsilon\frac{\mu^2}{r^3}\Big)
\dot{r}-r\dot{\phi}^{2}=0~,\\\\
r^2\dot{\phi}=h~.
\end{array}
\end{eqnarray}
Here the quantities $k$ and $h$ are constants that their physical meaning will be uncovered.
Without loss of generality, we restrict our attention to particles moving in the equatorial
plane, $\theta=\frac{\pi}{2}$. Replacing the second geodesic equation by a
first integral of the non-null geodesics with $g_{\mu\nu}\dot{x}^{\mu}
\dot{x}^{\nu}=c^2$,  we infer that the worldline $x(\tau)$ (where $\sigma=\tau$ is the proper time along the geodesics) of a massive
particle must satisfy the following equations
\begin{eqnarray}\label{e2-4*}
\begin{array}{ll}
\Big(1-\frac{2\mu}{r}+\varepsilon\frac{\mu^2}{r^2}\Big)\dot{t}=k~,\\\\
\Big(1-\frac{2\mu}{r}+\varepsilon\frac{\mu^2}{r^2}\Big)^{-1}\ddot{r}+\Big(1-\frac{2\mu}{r}
+\varepsilon\frac{\mu^2}{r^2}\Big)^2\Big(\frac{\mu}{r^2}-\varepsilon\frac{\mu^2}{r^3}\Big)
\dot{r}-r\dot{\phi}^{2}=c^2~,\\\\
r^2\dot{\phi}=h~.
\end{array}
\end{eqnarray}
In these equations a dot marks derivative with respect to the proper time $\tau$. Substituting the first
and third equations of (\ref{e2-4*}) into the second one, we obtain the deformed energy equation in the presence of a minimal measurable length as follows
\begin{equation}\label{e2-5}
\dot{r}^{2}+\frac{h^2}{r^2}\Big(1-\frac{2\mu}{r}+
\varepsilon\frac{\mu^2}{r^2}\Big)-\frac{2\mu c^2}{r}
+\varepsilon\frac{\mu^2 c^2}{r^2}=c^2(k^2-1)\,.
\end{equation}
This equation recovers the GR result in the limit of $\varepsilon\rightarrow0$.
We note that the right-hand side of Eq. (\ref{e2-5}) can be
interpreted as a constant of the motion so that in general case it must be $k=\frac{E}{E_0}$ (here $E$ and $E_0$ are respectively
the total and rest energies of the particle in its orbit). In order to
determine $r(\phi)$ which represents the shape of the orbit of rotating particles, we need another equation known as
the \emph{``shape equation''}. Since $\dot{r}$ in the above deformed energy equation can be written as
\begin{equation}\label{e2-6}
\frac{dr}{d\tau}=\frac{dr}{d\phi}~\frac{d\phi}{d\tau}=
\frac{h}{r^2}~\frac{dr}{d\phi}\,,
\end{equation}
by a change of variable as $u\equiv\frac{1}{r}$, we find
\begin{equation}\label{e2-7}
\frac{d^2u}{d\phi^2}+\Big(1+\varepsilon\frac{\mu^2c^2}{h^2}\Big)u
=\frac{\mu c^2}{h^2}+3\mu u^2-2\varepsilon\mu^2u^3~.
\end{equation}
This is our \emph{``minimal length deformed shape equation''}.
Setting conditions $\dot{r}=0=\ddot{r}$ ($u=\mbox{constant}$)
into Eq. (\ref{e2-7}), we obtain the following relation for the quantity $h$
\begin{equation}\label{e2-8}
h^2\equiv\frac{\mu c^2r^3-\varepsilon\mu ^2c^2r^2}{r^2-3\mu r
+2\varepsilon\mu^2}~.
\end{equation}
 Also Eqs. (\ref{e2-8}) and (\ref{e2-5}) jointly give the relation
\begin{equation}\label{e2-9}
k\equiv\sqrt{\frac{1-4\frac{\mu}{r}+(4+2\varepsilon)
\frac{\mu^2}{r^2}-\frac{4\varepsilon\mu ^3}{r^3}}{1-
3\frac{\mu}{r}+2\varepsilon\frac{\mu^2}{r^2}}}~,
\end{equation}
for the constant $k$. To determine whether the circular orbits are bounded in this setup,
we can use the relation $k=\frac{E}{E_0}$ which requires $E < E_0$. Therefore, the limits on
$r$ for the orbit to be bounded are given by $k = 1$,
which is satisfied if
\begin{equation}\label{e2-10}
4\mu(1-\frac{\varepsilon}{4})\leq r<\infty~.
\end{equation}
The net result of analysis done so far suggests that the closest bound circular orbit around
a black hole deformed by a natural cutoff as a minimum measurable length, can be formed at interval
$3\mu\leq r_{closest}\leq 5\mu$ which as compared with the prediction of GR, deviates with the value of $\mp|\varepsilon|\mu$.
Substituting the expression of (\ref{e2-8}) for $h^2$ into the third geodesic equation in (\ref{e2-4*}), we find
\begin{equation}\label{e2-11}
\dot{\phi}^{2}=\frac{\mu c^2r-\varepsilon
\mu^2c^2}{r^4-3\mu r^3+2\varepsilon\mu^2r^2}~.
\end{equation}
One can easily demonstrate that here $r=3\mu$ satisfies the geodesic equation ($\dot{\phi}^{2}>0$) while
in its GR counterpart this is not the case. In other words, unlike GR, based on the GUP deformed line element (\ref{e2-1}), for a free massive particle there
is the possibility of having a circular orbit at $r=3\mu$.\\
In which follows, we treat the stability of circular orbits in this minimal length deformed framework.
Using the deformed energy equation (\ref{e2-5}), the effective potential per unit mass can be identified as
\begin{equation}\label{e2-11}
U_{eff}(r)=-\frac{\mu c^2}{r}+\frac{h}{2}\Big(1+
\varepsilon\frac{\mu^2c^2}{2h^2}\Big)\frac{1}{r^2}
-\frac{h^2\mu}{r^3}+\varepsilon\frac{\mu^2h^2}{2r^4}~,
\end{equation}
which has an additional term proportional to $\frac{1}{r^4}$ as compared with the GR one. The additional term which depends on the sign of $\varepsilon$
can be thought as a repulsive/attractive term. A circular orbit occurs where $\frac{dU_{eff}(r)}{dr}=0$. So, differentiating Eq. (\ref{e2-11}), gives a cubic equation
as
\begin{equation}\label{e2-12}
\mu c^2 r^3-\Big(1+\varepsilon\frac{\mu^2c^2}{2h^2}\Big)
h^2r^2+3\mu h^2r-2\varepsilon \mu^2h^2=0~.
\end{equation}
This equation under the following conditions can results in one or three real solutions
\begin{equation}\label{e2-13}
\bar{h}<\sqrt{9-2\varepsilon}~~~~~\mbox{and}~~~~~
\bar{h}>\sqrt{9-2\varepsilon}~,
\end{equation}
respectively. Here, $\bar{h}\equiv\frac{h}{\mu c}$ is the dimensionless
angular momentum parameter. Corresponding to the first
and second conditions in (\ref{e2-13}), we extract the solutions for Eq. (\ref{e2-12}) as
\begin{equation}\label{e2-14}
\bar{r}_{0}\equiv \mu\Big[-2 \bar{h}\sqrt{1-\frac{\bar{h}^2}{9}-\frac{2\varepsilon}{9}}\sinh\Big(
\frac{1}{3}\sinh^{-1} \Big(\frac{\frac{-\bar{h}^3}{27}+(\frac{1}{2}-\varepsilon)\bar{h}^2-
\frac{\varepsilon \bar{h}}{2}+\frac{\varepsilon}{2}}{(1-\frac{\bar{h}^2}{9}-\frac{2\varepsilon}
{9})^{\frac{3}{2}}}\Big)\Big)\Big]
\end{equation}
and
\begin{eqnarray}\label{e2-15}
\begin{array}{ll}
\bar{r}_{1}\equiv \mu\Big[2 \bar{h}\sqrt{\frac{\bar{h}^2}{9}+\frac{2\varepsilon}{9}-1}\cos\Big(
\frac{1}{3}\cos^{-1} \Big(\frac{\frac{-\bar{h}^3}{27}+(\frac{1}{2}-\varepsilon)\bar{h}^2-\frac{\varepsilon \bar{h}}{2}+
\frac{\varepsilon}{2}}{
(\frac{\bar{h}^2}{9}+\frac{2\varepsilon}{9}-1)^{\frac{3}{2}}}\Big)\Big)+(\frac{\bar{h}}{3}+\frac{\varepsilon}{3\bar{h}})\Big]~,\\\\
\bar{r}_{2}\equiv \mu\Big[2\bar{h}\sqrt{\frac{\bar{h}^2}{9}+\frac{2\varepsilon}{9}-1}\cos\Big(
\frac{1}{3}\cos^{-1} \Big(\frac{\frac{-\bar{h}^3}{27}+(\frac{1}{2}-\varepsilon)\bar{h}^2-\frac{\varepsilon \bar{h}}{2}+
\frac{\varepsilon}{2}}{
(\frac{\bar{h}^2}{9}+\frac{2\varepsilon}{9}-1)^{\frac{3}{2}}}\Big)+\frac{2\pi}{3}\Big)+(\frac{\bar{h}}{3}+\frac{\varepsilon}{3\bar{h}})\Big]~,\\\\
\bar{r}_{3}\equiv \mu\Big[2\bar{h}\sqrt{\frac{\bar{h}^2}{9}+\frac{2\varepsilon}{9}-1}\cos\Big(
\frac{1}{3}\cos^{-1} \Big(\frac{\frac{-\bar{h}^3}{27}+(\frac{1}{2}-\varepsilon)\bar{h}^2-\frac{\varepsilon \bar{h}}{2}+
\frac{\varepsilon}{2}}{
(\frac{\bar{h}^2}{9}+\frac{2\varepsilon}{9}-1)^{\frac{3}{2}}}\Big)+\frac{4\pi}{3}\Big)+(\frac{\bar{h}}{3}+\frac{\varepsilon}{3\bar{h}})\Big]~,
\end{array}
\end{eqnarray}
respectively. Let us start our analysis from solution (\ref{e2-14}).
Physically, it means that there is only one extremum or turning point in the orbit for the relevant range
of $\bar{h}$. The solution is acceptable i.e. $\bar{r}_{0}>0$ only if $\frac{1}{2}\leq \varepsilon\leq1$
and $\sqrt{\frac{27}{10}}\leq\bar{h}<\sqrt{9-2\varepsilon}$.
Depending on the fixed values for $\varepsilon$ and $\bar{h}$, we will deal with different $\bar{r}_{0}$
limited to interval $0.01\mu<\bar{r}_{0}<5.3\mu$. Of course, some of $\bar{r}_{0}$'s  are excluded due to the violation of constraint
(\ref{e2-10}). So only the values belonging to the interval $3\mu\leq\bar{r}_{0}<5.3\mu$ are allowed. Putting (\ref{e2-14})
into the second derivative of the GUP-deformed effective potential
\begin{equation}\label{e2-16}
\frac{d^{2}U_{eff}}{dr^2}=3\mu c^2 r^2-2\mu^2 c^2
(\varepsilon+\bar{h}^2)r+3\mu^3\bar{h}^{2}~,
\end{equation}
we find that under all allowed values of $\bar{r}_{0}$, one has $\frac{d^{2}U_{eff}}{dr^2}|_{\bar{r}_{0}}>0$.
Therefore, it can be said that the solution (\ref{e2-14}) addresses \emph{exactly stable orbits} because the
local minimums in the potential are the locations of stable circular orbits.
At first look, it seems that (\ref{e2-15}) supports existence of three turning points. However, with a
direct analysis one can realize that this is not actually the case since the second and
third equations in (\ref{e2-15}) do not obey the condition $\bar{r}>0$ and therefore are unacceptable. Therefore,
here we face also with just one turning point as the former case. The first solution in (\ref{e2-15})
can be favorable if $-1\leq\varepsilon<0.3$ (except for $\varepsilon=0$) so that $\bar{r}_{1}>4.5\mu$.
As before, the solution also results in local minimums in the GUP-deformed effective potential, $\frac{d^{2}
U_{eff}}{dr^2}|_{\bar{r}_{1}}>0$. As a result, among four possible turning points only the couple $ \bar{r}_{0}$
and $\bar{r}_{1}$ have physical meaning so that both amazingly result in exactly stable circular orbits.
We can conceive ourselves with solutions $ \bar{r}_{0}$ and $\bar{r}_{1}$ as innermost and outermost orbits
respectively relevant to massive accretion disk around Schwarzschild black hole described by deformed metric (\ref{e2-1}).
Despite a feint similarity with GR one, the above results released in the presence of a minimal length
are different from what we have expected from GR-based effective potential. Here we face with a spectrum
of innermost orbits which the smallest orbit may be formed in $\bar{r}_{min}=3\mu$ while in GR for a
given value of $\bar{h}=2 \sqrt{3}$ we have an innermost circular orbit in definite value $r_{min}=6\mu$. On the other hand, the stability
status of the innermost circular orbit in GR is \emph{marginally}\footnote{
\emph{Generally, to have a circular orbit of massive particles under influence of a typical potential $U(r)$ at a radius $r=r_{0}$,
the conditions $U(r_{0})=0=U_{,r}(r_{0})$ should be fulfilled. In this case three behaviors are possible: a)~$U_{,rr}(r_{0})>0$: Massive
particles are on a stable circular orbit and are constrained to remain exactly at that value of $r_{0}$. b)~$U_{,rr}(r_{0})<0$:
The particles straddles the boundary between two regions with $U(r_{0})<0$; if the orbit is perturbed in one direction it falls into the hole,
while if it is perturbed in the opposite direction it moves outward and then turns back inward. In this case, the particles are on an unstable circular orbit.
c)~$U_{,rr}(r_{0})=0$: This is an intermediate case between the last two cases: a marginally stable circular orbit.}} \cite{Shapiro:2004, Bardeen:1972},
while here this orbit is exactly stable. In the context of GR, stability of circular motion of massive particles
on the innermost radius $r_{min}=6\mu$ is not lasting in response to a typical perturbation; rather it will collapse into the black hole. However, as we have seen, by considering a fundamental minimal length scale in the outer geometry of a Schwarzschild black hole, the stability status
of the innermost circular orbit improves towards an exactly stable orbit.

\section{The Redshift of the Black hole Accretion Disk}
As another important issue from a phenomenological viewpoint, here we study redshift of the photon gas accreting
on a none-rotational black hole which its outside geometry is determined by a minimal length deformed
Schwarzschild metric as (\ref{e2-1}). To avoid unnecessary complications, we take the disc to be oriented edge-on
to the observer so that all orbits are located in the plane of the observer and the disc (again we
set $\theta=\frac{\pi}{2}$ in our calculations). For the ratio of the photon's frequencies at the reception and emission points we find
\begin{equation}\label{e3-1}
\Delta_f=\frac{f_r}{f_e}=\frac{p_\mu(r)u^{\mu}_{r}}
{p_\mu(e)u^{\mu}_{e}}~,
\end{equation}
where $p_\mu(e)$ and $p_\mu(r)$ denote the photon four-momenta at emission and reception points,
respectively. Also $u^{\mu}_{e}$ and $u^{\mu}_{r}$ are the four-velocity of the particles at emission and reception, respectively.
The components of $u^{\mu}_{r}$ in the $(t,r,\theta,\phi)$ coordinates system with the assumption that the observer is fix at
infinity can be written as $u^{\mu}_{r}=(1,0,0,0)$. Besides, based on the assumption that the particles are moving in a circular orbit one has $u^{\mu}_{e}=
u^{0}_e(1,0,0,\Omega)$ so that

\begin{equation}\label{e3-2}
\Omega^2=(\frac{d\phi}{d\tau}~\frac{d\tau}{dt})^2~~~
\Rightarrow~~~\Omega^2=\frac{\Big(\mu c^2r-
\varepsilon\mu^2c^2\Big)\Big(1-\frac{2\mu}{r}+
\varepsilon\frac{\mu^2}{r^2}\Big)^2}{r^4-4\mu r^3
+4\mu^2(1+\frac{\varepsilon}{2})r^2-4\varepsilon\mu^3r}~.
\end{equation}
Now we can specify $u^{0}_{e}$ by using the fact that $g_{\mu\nu}u^{\mu}u^{\nu}= c^2$ for time-like geodesics. Under the condition that
the emitting particles are at a coordinate radius $r$, we then have
\begin{equation}\label{e3-3}
u^{0}_{e}\equiv \Big(1-\frac{2\mu}{r}+\varepsilon
\frac{\mu^2}{r^2}\Big)^{-1}\Big[\Big(1-\frac{2\mu}{r}
+\varepsilon\frac{\mu^2}{r^2}\Big)^{-1}-
\frac{\Big(\mu c^2r^2-\varepsilon\mu^2c^2r\Big)}{r^3
-4\mu r^2+4\mu^2(1+\frac{\varepsilon}{2})r-4\varepsilon\mu^3}
\Big]^{-\frac{1}{2}}~.
\end{equation}
Therefore we can rewrite the general expression for
(\ref{e3-1}) as
\begin{equation}\label{e3-4}
\Delta_f=\frac{p_0(r)}{p_0(e)u^{0}_{e}+p_3(e)u^{3}_{e}}=
\frac{1}{u^{0}_e}\Big(1\pm\frac{p_3(e)}{p_0(e)}\Omega\Big)
^{-1}~.
\end{equation}
The plus sign in this equation corresponds to the emitting photons on the side of the disc moving towards the observer,
while the minus sign corresponds to the photons on the other side. To fix the ratio $\frac{p_3(e)}{p_0(e)}$ we
can apply the null geodesic relation $g^{\mu\nu}p_{\mu}p_\nu=0$ for the photon's worldline as
\begin{equation}\label{e3-5}
\frac{1}{c^2}\Big(1-\frac{2\mu}{r}+\varepsilon\frac{\mu^2}{r^2}
\Big)^{-1}p_0^{2}-\Big(1-\frac{2\mu}{r}+\varepsilon\frac{\mu^2}
{r^2}\Big)p_1^{2}-\frac{1}{r^2}p_3^{2}=0~.
\end{equation}
A special and simple case happens when the photon is emitted from matter moving transversely to the observer i.e. $\phi=0$
or $\phi=\pi$ which results in $p_3(e)=0$. We note that in this case the disc is viewed face-on. Accordingly, the
observed frequency ratio in the presence of a fundamental minimal length is given by
\begin{equation}\label{e3-6}
\Delta_f\equiv
\sqrt{\Big(1-\frac{2}{\tilde{r}}+\frac{\varepsilon}
{\tilde{r}^{2}}\Big)-\frac{(\tilde{r}^2-\varepsilon
\tilde{r})\Big(1-\frac{2}{\tilde{r}}+\frac{\varepsilon}
{\tilde{r}^{2}}\Big)^2}{\tilde{r}^3-4\tilde{r}^2+(4+
2\varepsilon)\tilde{r}-4\varepsilon}}~,
\end{equation}
where $\tilde{r}\equiv\frac{r}{\mu}$ is a dimensionless parameter. It is easily verifiable that in
the absence of the minimal length modification (i.e. for $\varepsilon\rightarrow0$), equation (\ref{e3-6})
recovers its GR counterpart as
\begin{equation}\label{e3-6*}
\Delta_f\equiv \sqrt{1-\frac{3}{\tilde{r}}}~.
\end{equation}
The other special case occurs when the matter is moving either directly towards or away from the observer i.e.
$\phi=\pm\frac{\pi}{2}$ (here the disc is viewed edge-on). In this case after a little calculation, one obtains
the following relation
\begin{equation}\label{e3-7}
\Delta_f\equiv
\frac{\sqrt{\Big(1-\frac{2}{\tilde{r}}+\frac{\varepsilon}
{\tilde{r}^{2}}\Big)-\frac{(\tilde{r}^2-\varepsilon\tilde{r})
\Big(1-\frac{2}{\tilde{r}}+\frac{\varepsilon}{\tilde{r}^{2}}
\Big)^2}{\tilde{r}^3-4\tilde{r}^2+(4+2\varepsilon)\tilde{r}-
4\varepsilon}}}{1\pm\Big(1-\frac{2}{\tilde{r}}+\frac{\varepsilon}
{\tilde{r}^{2}}\Big)\sqrt{\frac{(\tilde{r}^2-\varepsilon\tilde{r})}
{\tilde{r}^3-4\tilde{r}^2+(4+2\varepsilon)\tilde{r}-4\varepsilon}}}~,
\end{equation}
for the photon frequency shift in the context of GUP. Once again, by discarding the minimal length effects in our calculations, Eq. (\ref{e3-7}) recovers the same
expression that we expect from GR
\begin{equation}\label{e3-8}
\Delta_f\equiv \frac{(1-\frac{3}{\tilde{r}})^{\frac{1}{2}}}
{1\pm(\tilde{r}-2)
^{-\frac{1}{2}}}~.
\end{equation}
\begin{figure}
\begin{center}
\begin{tabular}{c}\hspace{-1cm}\epsfig{figure=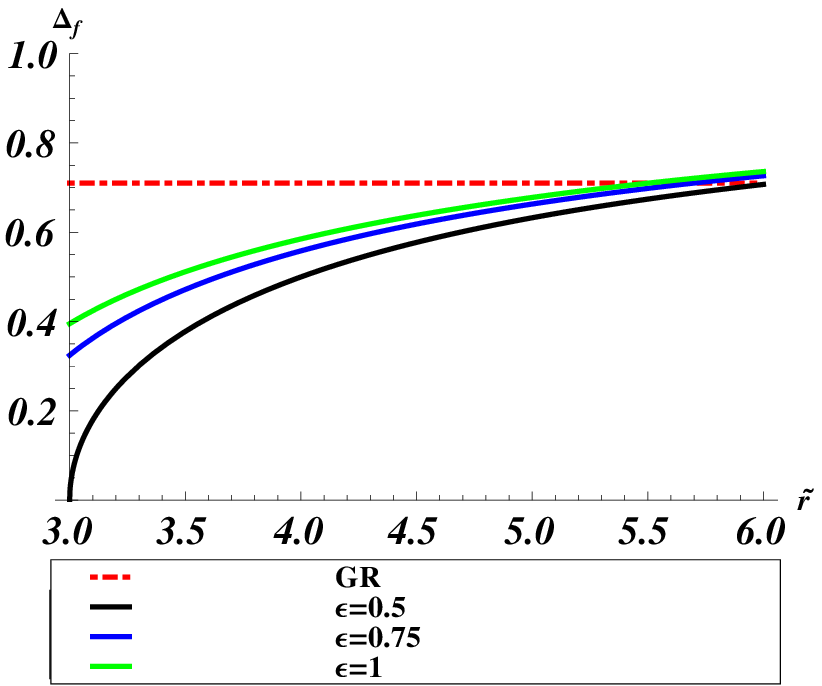,width=5.8cm}
\hspace{1cm} \epsfig{figure=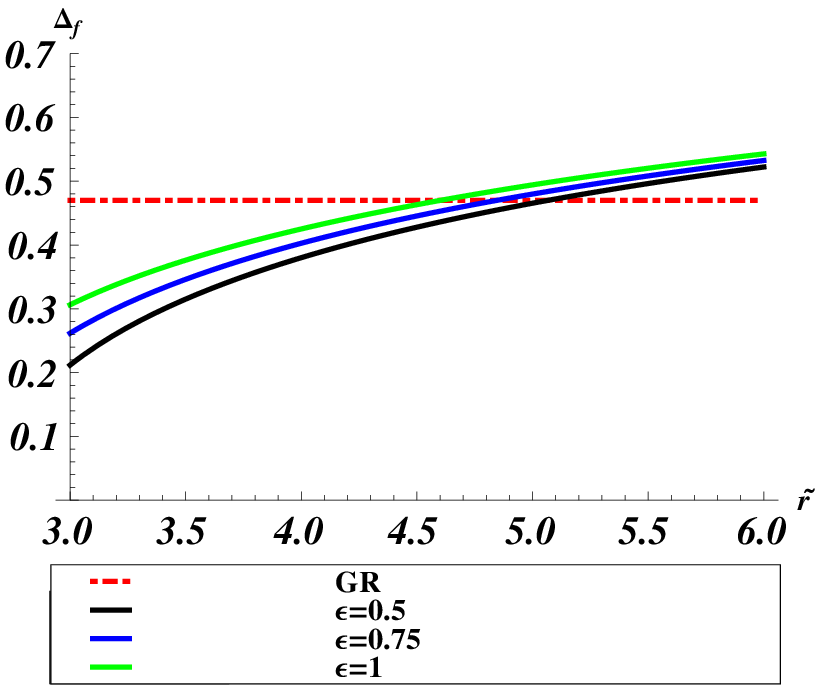,width=5.8cm}
\hspace{1cm} \epsfig{figure=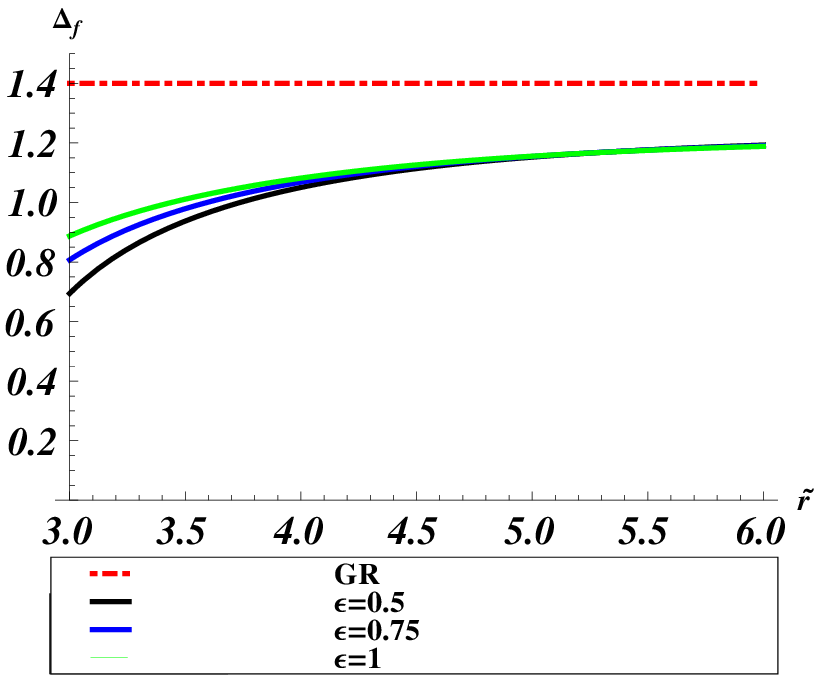,width=5.8cm}
\end{tabular}
\end{center}
\caption{\footnotesize {\emph{Photon frequency
shift given by Eqs. (\ref{e3-6}) (left panel),
(\ref{e3-7}) with positive sign (middle panel)
and (\ref{e3-7}) with negative sign (right panel)
as a function of the dimensionless parameter $\tilde{r}$
for some values of $\varepsilon$. Red dot-dashed
horizontal line in each three panels corresponds to the smallest values
that one expects from GR by substituting $\tilde{r}_{min}=6$
into Eqs. (\ref{e3-6*}) and (\ref{e3-8}) with positive sign and
in (\ref{e3-8}) with negative sign, respectively}. }}
\label{fig:1}
\end{figure}
In Fig. 1, we have depicted the qualitative behavior of Eqs. (\ref{e3-6}) and (\ref{e3-7}) to see the
impact of the minimal length deformation on GR prediction. In contrast to the standard accretion disk formalism,
here there is a possibility of radiation for regions smaller than $\tilde{r}_{min}=6$. The plots depicted in
Fig. 1 implicitly display the fact that the smallest frequency shifts in the presence
of a minimal measurable length are below the GR prediction. This point is reliable since in the standard GR
when one computes the radiation of a standard accretion disk by taking rotation of the
black hole (Kerr metric) into account, the smallest frequency shifts then could be lower than the case without rotation \cite{Shapiro:2004}.

\section{Gravitational Tidal Forces Around Black Hole }

In the same streamline as previous sections, now we are going to study gravitational tidal forces near a GUP-deformed
Schwarzschild black hole described by the deformed line element (\ref{e2-1}). A set of orthonormal basis vectors defining an instantaneous inertial
rest frame for one of the particles can be considered as
\begin{equation}\label{e4-1}
(\hat{e}_{t})^\alpha=\frac{1}{c}\Big(1-\frac{2\mu}{r}+\frac{\varepsilon\mu^2}
{r^{2}}\Big)^{-\frac{1}{2}}\delta_{t}^{\alpha},~~~(\hat{e}_{r})^\alpha=
\Big(1-\frac{2\mu}{r}+\frac{\varepsilon\mu^2}
{r^{2}}\Big)^{\frac{1}
{2}}\delta_{r}^{\alpha},~~~(\hat{e}_{\theta})^\alpha=\frac{1}{r}\delta_{\theta}^
{\alpha},~~~(\hat{e}_{\phi})^\alpha=\frac{1}{r\sin\theta}\delta_{\phi}^
{\alpha}~,
\end{equation}
where $\alpha$ runs from $0$ to $3$ correspond to
$(t,r,\theta,\phi)$. The general expression for calculation of the tidal forces is written as
\begin{equation}\label{e4-2}
\frac{d^2\xi^{\alpha}}{d\tau^2}=c^2R^{\hat{\alpha}}_{\,\,\hat{0}\hat{0}\hat{\gamma}}
\xi^{\hat{\gamma}},~~~~R^{\hat{\alpha}}_{\,\,\hat{0}\hat{0}\hat{\gamma}}=R^{\mu}_{\,\,\sigma\nu\rho}
(\hat{e}^\alpha)_{\mu}(\hat{e}_{\beta})^{\sigma}(\hat{e}_{\lambda})^{\nu}(\hat{e}_{\delta})^{\rho}~.
\end{equation}
Here $R^{\mu}_{\,\,\sigma\nu\rho}$ denotes the Riemann tensor in the Schwarzschild coordinates with the following general expression
\begin{equation}\label{e4-3}
R^{\mu}_{\,\,\sigma\nu\rho}=\partial_\nu\Gamma^{\mu}_{\,\,\sigma\rho}-\partial_\rho\Gamma^{\mu}_{\,\,\sigma\nu}
+\Gamma^{e}_{\,\,\sigma\rho}\Gamma^{\mu}_{\,\,e\nu}-\Gamma^{e}_{\,\,\sigma\nu}\Gamma^{\mu}_{\,\,e\rho}~,
\end{equation}
where
\begin{equation}\label{e4-4}
\Gamma^{\alpha}_{\,\,\beta\gamma}=\frac{1}{2}g^{\alpha\zeta}\Big(
\partial_{\beta} g_{\zeta\gamma}+\partial_{\gamma} g_{\beta \zeta}-
\partial_{\zeta} g_{\beta\gamma}\Big)~.
\end{equation}
To receive our goal in this section, we need to know the minimal length
deformed expressions of the Riemann tensor. The non-zero elements of
the metric components $g_{\alpha\beta}$ from the GUP-deformed line element
(\ref{e2-1}) are given by
\begin{equation}\label{e4-5}
g_{tt}=c\Big(1-\frac{2\mu}{r}+\frac{\varepsilon\mu^2}
{r^{2}}\Big),~~~g_{rr}=-\Big(1-\frac{2\mu}{r}+\frac{\varepsilon\mu^2}
{r^{2}}\Big)^{-1},~~~g_{\theta\theta}=-r^2,~~~~g_{\phi\phi}=-r^2\sin^{2}\theta~.
\end{equation}
Then, the non-zero connection coefficients are calculated as
\begin{eqnarray}\label{e4-6}
\begin{array}{ll}
\Gamma^{t}_{\,\,tr}=\Gamma^{t}_{\,\,rt}=\Big(1-\frac{2\mu}{r}+\frac{\varepsilon\mu^2}
{r^{2}}\Big)^{-1}\Big(\frac{2\mu}{r^2}-\frac{2\varepsilon\mu^2}
{r^{3}}\Big)~,\\\\
\Gamma^{r}_{\,\,tt}=-\Gamma^{r}_{\,\,rr}=\Big(1-\frac{2\mu}{r}+\frac{\varepsilon\mu^2}
{r^{2}}\Big)\Big(\frac{\mu}{r^2}-\frac{\varepsilon\mu^2}
{r^{3}}\Big)~,\\\\
\Gamma^{r}_{\,\,\phi\phi}=-r\Big(1-\frac{2\mu}{r}+\frac{\varepsilon\mu^2}
{r^{2}}\Big)~,\\\\
\Gamma^{\theta}_{\,\,r\theta}=\Gamma^{\theta}_{\,\,\theta r}=\Gamma^{\phi}_{\,\,r\phi}=
\Gamma^{\phi}_{\,\,\phi r}=\frac{1}{r}\,.
\end{array}
\end{eqnarray}
Now we can split Eq. (\ref{e4-2}) to the following equations
\begin{eqnarray}\label{e4-6*}
\begin{array}{ll}
\frac{d^2\xi^{\hat{r}}}{d\tau^2}=\Big(1-\frac{2\mu}{r}+\frac{\varepsilon\mu^2}
{r^{2}}\Big)^{-\frac{3}{2}}\Big[R^{r}_{\,\,ttt}(\hat{e}_{\gamma})^{t}+R^{r}_{\,\,ttr}(\hat{e}_{\gamma})
^{r}+R^{r}_{\,\,tt\theta}(\hat{e}_{\gamma})^{\theta}+
R^{r}_{\,\,tt\phi}(\hat{e}_{\gamma})^{\phi}\Big]\xi^{\hat{\gamma}}~,\\\\
\frac{d^2\xi^{\hat{\theta}}}{d\tau^2}=r\Big(1-\frac{2\mu}{r}+\frac{\varepsilon\mu^2}
{r^{2}}\Big)^{-1}\Big[R^{\theta}_{\,\,ttt}(\hat{e}_{\gamma})^{t}+R^{\theta}_{\,\,ttr}(\hat{e}_{\gamma})
^{r}+R^{\theta}_{\,\,tt\theta}(\hat{e}_{\gamma})^{\theta}+
R^{\theta}_{\,\,tt\phi}(\hat{e}_{\gamma})^{\phi}\Big]\xi^{\hat{\gamma}}~,\\\\
\frac{d^2\xi^{\hat{\phi}}}{d\tau^2}=r\Big(1-\frac{2\mu}{r}+\frac{\varepsilon\mu^2}
{r^{2}}\Big)^{-1}\Big[R^{\phi}_{\,\,ttt}(\hat{e}_{\gamma})^{t}+R^{\phi}_{\,\,ttr}(\hat{e}_{\gamma})
^{r}+R^{\phi}_{\,\,tt\theta}(\hat{e}_{\gamma})^{\theta}+
R^{\phi}_{\,\,tt\phi}(\hat{e}_{\gamma})^{\phi}\Big]\xi^{\hat{\gamma}}~,
\end{array}
\end{eqnarray}
respectively. By putting the connection coefficients (\ref{e4-6}) into these relations
along with using coordinate basis (\ref{e4-1}), after a little algebra one finally
arrives at the following deformed expressions
\begin{equation}\label{e4-7}
\frac{d^2\xi^{\hat{r}}}{d\tau^2}=\Big(\frac{2\mu c^2}{r^3}-\frac{3\varepsilon \mu^2 c^2}{r^4}
\Big)\xi^{\hat{r}},~~~~\frac{d^2\xi^{\hat{\theta}}}{d\tau^2}=-\Big(\frac{\mu c^2}{r^3}-
\frac{\varepsilon \mu^2 c^2}{r^4}\Big)\xi^{\hat{\theta}},~~~~\frac{d^2\xi^{\hat{\phi}}}
{d\tau^2}=-\Big(\frac{\mu c^2}{r^3}-\frac{\varepsilon \mu^2 c^2}{r^4}\Big)\xi^{\hat{\phi}}~,
\end{equation}
correspond to the spatial components of the gravitational tidal forces between two particles around a Schwarzschild black hole
with geometry deformed by a fundamental minimal length. First of all, we see that in the limit $\varepsilon\rightarrow0$,
the above expressions recover their GR counterparts. At once, one finds that radial dependence of the GUP deformation term in each three
components is the same and as $\frac{1}{r^4}$, while the GR term is as $\frac{1}{r^3}$. When the GUP-deformed gravitational tidal force
becomes dominant, we are faced with more different possibilities in comparison with standard GR. If we take the positive sign for $\varepsilon$
($0<\varepsilon\leq1$), the GUP-deformation term then resists against tension or stretching generated by GR term in the radial direction
$ \xi^{\hat{r}}$ as far as $r\leq\frac{3}{2}\mu$ where transition from tension to compression happens. Also GUP-deformation
term creates a repulsion against pressure or compression arising from GR term in the transverse directions ($ \xi^{\hat{\theta}}$
and $ \xi^{\hat{\phi}}$) as far as $r\leq\mu$ where compression converts into tension. On the other hand, deformed line element
(\ref{e2-1}) discloses this point that in the presence of a minimal length, the Schwarzschild radius cannot be fixed
at the value $r=2\mu$ rather, depending on $0<\varepsilon\leq1$, it is located in the range $\frac{3}{2}\mu\leq r<2\mu$.
As a result, a freely falling particle passing the smallest Schwarzschild radius and entering the region $\mu<r<\frac{3}{2}\mu$,
gets squeezed in all directions. However, as it arrives at the region $0<r<\mu$, despite contraction in the
radial direction continues, the particle expands in the transverse directions. We note that for possible negative sign of $\varepsilon$, the effect of tension and compression enhance
in the radial and transverse directions, respectively. Unlike the former possibility, in this case as GR, no transition takes place.

\section{The Geodetic Drift Rate}

Inspired by the fact that the motion of spin vector of a test body (we mean a very small
object with spin such as a small gyroscope) can be employed
to probe geometry of a curved spacetime, here we demand this issue for the GUP-deformed Schwarzschild
geometry (\ref{e2-1}) to study geodetic  drift rate. Due to the motion of a test body along a
timelike geodesic, four-velocity $u(\tau)$ is parallel-transported along its worldline. Therefore, in some coordinate
system, the components of $u(\tau)$ satisfy the following equation
\begin{equation}\label{e5-1}
\frac{du^{\mu}}{d\tau}+\Gamma^
{\mu}_{\,\,\nu\sigma}u^{\nu}
u^{\sigma}=0~.
\end{equation}
Assuming that the spin of the test body is characterized by the $s(\tau)$ along the geodesic, one can require the
following orthogonality condition
\begin{equation}\label{e5-2}
s.u=g_{\mu\nu}u^{\mu}u^{\nu}~,
\end{equation}
at all points along the geodesic. This condition is imposed from the fact that the spin vector has no timelike component in the instantaneous rest
frame of the test body. Parallel transport of $u(\tau)$ along its geodesic dictates
\begin{equation}\label{e5-3}
\frac{ds^{\mu}}{d\tau}+\Gamma^
{\mu}_{\,\,\nu\sigma}s^{\nu}
u^{\sigma}=0~.
\end{equation}
Now by supposing that the test body is in a circular orbit of coordinate radius $r$ in the equatorial plane $(\theta=\frac{\pi}{2}$) of the Schwarzschild geometry deformed
by a minimal length, Eq. (\ref{e5-3}) then reduces to
\begin{eqnarray}\label{e5-4}
\begin{array}{ll}
\frac{d s^{t}}{d\tau}+\Gamma^{t}_{\,\,rt}s^ru^t=0~,\\\\
\frac{d s^{r}}{d\tau}+\Gamma^{r}_{\,\,tt}s^tu^t+\Gamma^{r}_
{\,\,\phi\phi}s^\phi
u^\phi=0~,\\\\
\frac{d s^{\theta}}{d\tau}=0~,\\\\
\frac{d s^{\phi}}{d\tau}+\Gamma^{\phi}_{\,\,r\phi}
s^ru^\phi=0~.
\end{array}
\end{eqnarray}
The four-velocity of the test body can be written as $u^{\mu}=u^t(1,0,0,\Omega)$ (see Section 2). Therefore, from Eq. (\ref{e5-2}) one can extract the following result
\begin{equation}\label{e5-5}
s^{t}=\frac{\Omega r^2}{(1-\frac{2\mu}{r}+
\varepsilon\frac{\mu^2}{r^2})}s^{\phi}~.
\end{equation}
Using this relation along with connection coefficients listed in (\ref{e4-6}), one can rewrite Eq.(\ref{e5-4})
as
\begin{eqnarray}\label{e5-6}
\begin{array}{ll}
\frac{d s^{\phi}}{d\tau}+\frac{c^2}{\Omega}\Big(\frac{\mu}{r^4}-
\varepsilon\frac{\mu^2}{r^5}\Big)u^t s^r=0~,\\\\
\frac{d s^{r}}{d\tau}+\Omega\bigg(\Big(\frac{\mu}{c^2}
-\varepsilon\frac{\mu^2}{c^2r}\Big)-
(r-2\mu+\varepsilon\frac
{\mu^2}{r})\bigg)u^ts^\phi=0~,\\\\
\frac{d s^{\theta}}{d\tau}=0~,\\\\
\frac{d s^{\phi}}{d\tau}+\frac{\Omega}{r}u^ts^r=0~.
\end{array}
\end{eqnarray}
It is straightforward to prove that in the absence of $\varepsilon$, the above four independent equations reduce to three equations since the
first and fourth ones are actually the same in this case. At this point one usually converts the $\tau$-derivatives in the above equations to $t$-derivatives which will
be done through the relation $u^t=\frac{dt}{d\tau}$. Therefore, except for the second equation, the rest become as
\begin{eqnarray}\label{e5-6*}
\begin{array}{ll}
\frac{d s^{\phi}}{dt}+\frac{1}{\Omega}\Big(\frac{\mu c^2}{r^4}-
\varepsilon\frac{\mu^2 c^2}{r^5}\Big)s^r,~~~~ \frac{d s^{\theta}}{dt}=0,~~~~
\frac{d s^{\phi}}{dt}+\frac{\Omega}{r}s^r=0~,
\end{array}
\end{eqnarray}
respectively. With these relations, the second equation in (\ref{e5-6}) can be rewritten as the following two forms
\begin{eqnarray}\label{e5-7}
\begin{array}{ll}
\frac{d^2 s^{r}}{dt^2}+\Omega_{1}^{2}s^r=0,~~~~~~\Omega_{1}\equiv\sqrt{\Big
(\frac{\mu c^2}{r^3}-\frac{3\mu^2 c^2}{r^4}\Big)+\varepsilon
\big(\frac{\mu^3 c^2}{r^5}-\frac{\mu^2 c^2}{r^4}\big)}~,\\\\
\frac{d^2 s^{r}}{dt^2}+\Omega_{2}^{2}s^r=0,~~~~~\Omega_{2}\equiv\Omega\sqrt{1-
\frac{3\mu}{r}+2\varepsilon\frac{\mu^2}{r^2}}
\end{array}
\end{eqnarray}
in terms of $t$-derivative respectively. Finally, with assumption that the initial spatial direction $\vec{s}$ of the spin vector to
be radial (i.e. $s^\theta(0)=s^{\phi}(0)=0$), the corresponding solutions to the system of equations (\ref{e5-6*}) and (\ref{e5-7})
then are obtained as
\begin{eqnarray}\label{e5-8}
\begin{array}{ll}
s^{\phi}(t)=-\frac{\Omega_3}{\Omega_1}s^{r}(0)\sin\Omega_1 t~,~~~~~~
\Omega_3\equiv \Big(\frac{\mu}{\Omega r^4}-\varepsilon\frac{\mu^2}
{\Omega r^5}\Big)~,\\\\
s^{\theta}(t)=0~,\\\\
s^{\phi}(t)=-\frac{\Omega}{r\Omega_2}s^{r}(0)\sin\Omega_2 t~.
\end{array}
\end{eqnarray}
and
\begin{equation}\label{e5-9}
s^{r}(t)=s^{r}(0)\cos \Omega_1 t~,~~~~~~s^{r}(t)=s^{r}(0)\cos
\Omega_2t
\end{equation}
respectively. Once again we stress that by suppressing the effect of the minimal length, the first
and the third cases in Eq. (\ref{e5-8}) are the same so that two equations in (\ref{e5-9}) reduce to one equation, as we
expect from GR. The solution (\ref{e5-8}) reveals clearly that the angular parts of the spin vector rotate relative to the radial direction with two distinct
angular speeds $\Omega_{1,\,2}$ in the negative $\phi$-direction. On the other hand, the radial direction
itself rotates with the same angular speeds, this time in the positive $\phi$-direction. The difference
between these two angular speeds leads to a phenomena known as \emph{`` geodesic precession effect"} or
\emph{``geodetic drift rate"}. This phenomena is detectable via angle \footnote{\emph{After one complete
revolution which takes coordinate time $t=\frac{2\pi}
{\Omega}$, the final direction of spin vector does not return to the starting state, rather there would be a small shift i.e.
$2\pi+\varphi$. }} $\varphi_{1,\,2}=\frac{2\pi}{\Omega}
(\Omega- \Omega_{1,\,2})$, i.e
\begin{equation}\label{e5-10}
\varphi_1=2\pi\Bigg(1-\sqrt{\frac{\Big((\tilde{r}-3)+\varepsilon(\frac{1}
{\tilde{r}}-1)\Big)\Big(1-4\tilde{r}+(4+2\varepsilon)\tilde{r}^2-4\varepsilon
\tilde{r}^3\Big)}{(\tilde{r}-\varepsilon)\Big(1-2\tilde{r}+\varepsilon\tilde{r}
^2\Big)^2}}\Bigg),~~~~~ \varphi_2=2\pi \Big(1-\sqrt{1-\frac{3}{\tilde{r}}+
\varepsilon\frac{2}{\tilde{r}^2}}\Big)~.
\end{equation}
With a simple calculation it can be shown that in the limit $\varepsilon\rightarrow0$, we have
$\varphi_1=\varphi_2=2\pi\Big(1- \sqrt{1-\frac{3}{\tilde{r}}}\Big)$, as expected from GR. We stress
that $\varphi_1$ addresses the geodesic precession effect only for the positive values of $\varepsilon$,
exclusively in the region $\tilde{r}<3$ which is forbidden from GR perspective. However,
it cannot be valid physically because it diverges (see Fig. 2 (left panel)).
\begin{figure}
\begin{center}
\begin{tabular}{c}\hspace{-1cm}\epsfig{figure=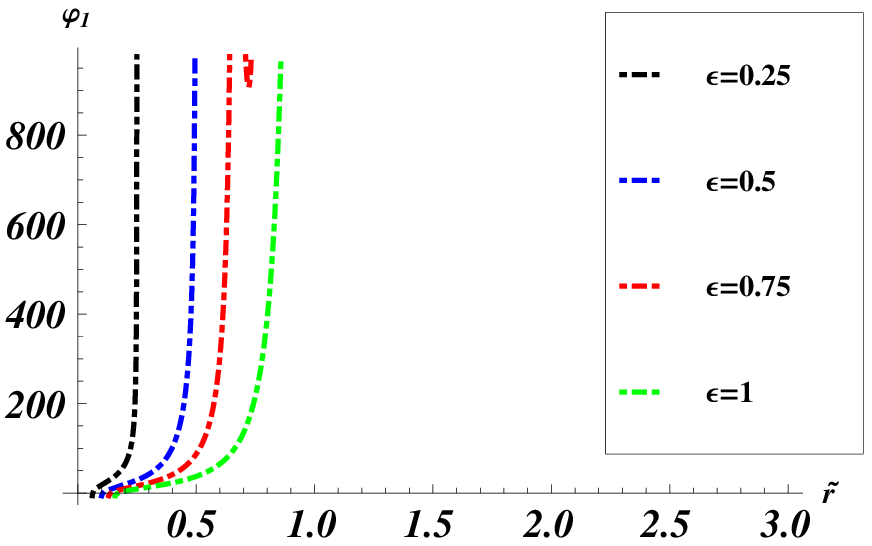, width=7.5cm}
\hspace{1cm} \epsfig{figure=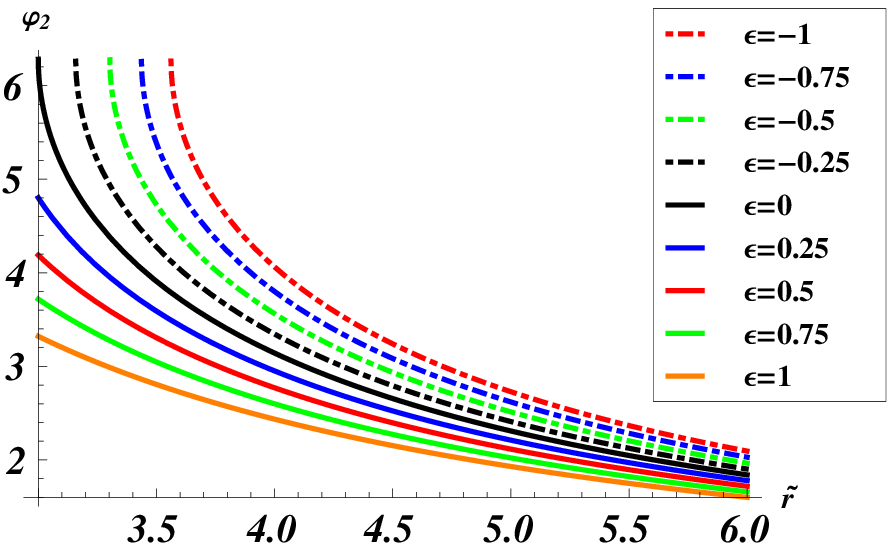,width=7.5cm}
\end{tabular}
\end{center}
\caption{\footnotesize {\emph{Behavior of $\varphi_1$
(left panel) and $\varphi_2$ (right panel) given by Eq.
(\ref{e5-10}) as a function of the dimensionless parameter
$\tilde{r}$ for different values of $\varepsilon$.}}}
\label{fig:1}
\end{figure}
Nevertheless, qualitative behavior of $\varphi_2$ in Fig. 2 (right panel) is noteworthy because in
the same allowable range in GR (that is, $\tilde{r}>3$) it shows the effect of GUP-deformation on the geodetic
drift rate. In other words, for $0<\varepsilon\leq1$, $\varphi_2$ shifts towards below the GR result while
for $-1\leq\varepsilon<0$ it shifts up to the GR result. Although this effect is very tiny, it may be
detectable experimentally by measuring the spacelike spin vector of a gyroscope in an
orbiting spacecraft. One of the main goals of the \emph{Gravity Probe B} project as a space-based
experiment\footnote{\emph{This project launched from April 2004 by NASA and data have been collected
from August 2004 up to August 2005 via four cryogenic gyroscopes in earth orbit.}}, was test of the geodetic
drift rate predicted by GR \cite{GP:2011}. After analyzing all data, the value has reported is nearly $6.6018$
arc-second/Yr (or $32\times10^{-6}$ Rad/Yr) which has a negligible difference with the prediction of GR,i.e.
nearly $8$ arc-second/Yr (or $38\times10^{-6}$ Rad/Yr). We remind that within the context of GR,
this deviation can be explained by the fact that the actual geometry outside of rotating Earth is Kerr geometry not the static Schwarzschild one. For a more detailed discussion
we provide a numerical analysis on the dependency of $\varphi_2$ to $\varepsilon$ (see Table I).
The content of Table I explicitly reflects the fact that incorporating a fundamental
minimal length scale into the Schwarzschild geometry with the relevant positive dimensionless parameter
constrained as $0<\varepsilon<0.14$, leads to improving the prediction of GR for a better compliance with experiment.
As is shown in Table I, by fixing $\varepsilon=0.1374$ we arrive at $\varphi_2=6.6020$ arc-second/Yr
which is closer to released value from \emph{Gravity Probe B} in comparison with \emph{GR+Kerr metric} one.

It is worth to note that a similar situation has occurred
in section 3 about the issue of \emph{``the smallest frequency shifts''}. In the light of the results obtained in this paper, we are able to
justify at least a part of the small deviation between the GR and experimental results by incorporation of the quantum gravity effect
via a minimal measurable length into the Schwarzschild geometry without considering the small effect of the Earth rotation.
The attractive point of this analysis is that a simple deformation of the Schwarzschild metric as (\ref{e2-1}) has the potential to derive more or less the same phenomenological adjustment arising from a complicated metric such as the Kerr metric.
\begin{table}[t]
\caption{\emph{The relevant values of angel $\varphi_2$ in terms of the dimensionless
parameter $\varepsilon$ by fixed $\tilde{r}=3$ for the nearest Earth's orbit as expected
from GR. By increasing $\varepsilon$ the value of $\varphi_2$ is improving towards
\textbf{Gravity Probe B} result}. }
\begin{tabular}{|c|c|}
  \hline
  Dimensionless Parameter $\varepsilon$& Angel $\varphi_2$ (arc-second/Yr)\\
  \hline
  0 & 8 \\ \hline
  0.02 & 7.4666 \\ \hline
  0.04 & 7.2456 \\ \hline
  0.06 & 7.0761 \\ \hline
  0.08 & 6.9332\\ \hline
  0.1 & 6.8073 \\ \hline
  0.12 & 6.6935\\  \hline
  \textbf{0.1374} & \textbf{6.6020}\\  \hline \hline
  Angel $\varphi_2$  released from \emph{Gravity Probe B}
  & Angel $\varphi_2$ predicted by \emph{GR+ Kerr metric}\\  \hline
  \textbf{$6.6018\pm0.0183$} arc-second/Yr \cite{GP:2011}& \textbf{$6.6061$} arc-second/Yr
  \cite{GP:2011}\\ \hline
  \end{tabular}
\end{table}

\section{Summary and Conclusions}

The authors in Ref. \cite{Fabio:2015}, without violation of the equivalence principle and geodesic equation,
have presented an appropriate deformation of the Schwarzschild metric (\ref{e2-1}) which results
in retrieval of the Hawking temperature drawn out from the Gravitational Uncertainty Principle
(GUP). In more details, the deformation parameter $\varepsilon$ in the line element (\ref{e2-1})
can be linked with GUP deformation parameter $\beta$ which addresses the existence of a minimal
measurable length in nature. The deformed metric (\ref{e2-1}) addresses a Schwarzschild geometry which is equipped with a natural
cutoff as a minimal measurable length. In recent years, there has been a lot of attention on quantum gravity
phenomenology by focusing on predictions indictable at low energies accessible in current our future experiments. In this
paper, by considering four important issues prevalent in astrophysical systems such as a Schwarzschild black hole, we
treated some phenomenological consequences of QG deformed Schwarzschild metric (\ref{e2-1})
hoping to shed light on general relativity results, their possible modification and also confrontation with experiments. The summary and main results are as follows:
\begin{itemize}
  \item As the first issue, by considering a minimal length in our calculation, we derived the deformed
  energy equation for the $r$-coordinate along with relevant shape equation to explore stability
  of circular trajectories of massive particles (particularly, accretion disk around a Schwarzschild black hole which is deformed by a
  minimal length). We firstly found that by fixing the deformation parameter into interval
  $-1\leq\varepsilon\leq1$ (except for $\varepsilon=0$), unlike our expectation from GR, the closest
  bound circular orbit may be formed at range $3\mu\leq r_{closet}\leq5\mu$. As a direct consequence of incorporating the minimal length
  into Schwarzschild geometry, we showed that a free massive particle is able to keep its
  circular orbit at $r=3\mu$, which is impossible in GR. Then by regarding the minimal length into the effective
  potential as well as applying some explicit constraints on dimensionless angular parameter
  $\bar{h}$ and $\varepsilon$, we extracted two physical solutions among four possible ones
  which can be thought as innermost and outermost circular orbits in this setup. However, these two solutions
  have a different status than what are derived from GR-based effective potential. One will
  find that here depending on $\varepsilon$ and $\bar{h}$, there is a spectrum of innermost
  orbits which the smallest orbit is formed at $\tilde{r}_{min} = 3\mu$. This is the case in the situation that
  in the context of GR for a given value of $\bar{h}=2\sqrt{3}$ we have only one innermost circular orbit at $r_{min}=6\mu$.
  More importantly, while the stability status of the innermost circular orbit in GR is
  marginally, here we obviously showed that it is exactly stable. Namely, according to GR
  despite the stable circular motion of massive particles on the innermost radius
  $r_{min} = 6\mu$, it is not durable against a typical perturbation so that
  it falls down into the black hole. Overall, our analysis reveals that embedding a
  fundamental minimal length into the outer geometry of the Schwarzschild black
  hole improves the stability status of both mentioned circular orbits towards exactly
  stable one. It is interesting to note that,
  the shape of the metric (1) signals obviously a fascinating resemblance with the Reissner-Nordstr\"{o}m (RN) metric. RN metric
  is a solution of the Einstein-Maxwell equations which points out a non-rotating
  charged black hole with the gravitational mass $m$ and the electric charge $e$ so that
  $\varepsilon\equiv\frac{e^2}{m^2}$. Focusing on the underlying issues
  within the NR metric, one explores that the behavior of massive accretion discs
  highly depends on the charge to mass ratio $\frac{e}{m}$, see for instance \cite{Rajab:2010}.
  On the other hand, similar to what happened in our case, for a NR black hole also
  there is a continuous region of stability along spacelike geodesics from the innermost to
  outermost circular orbits, as shown in \cite{Ruffini:2011}.

  \item As the second issue, we have calculated in some straightforward cases the redshift of the photon
  gas accretion around a GUP-deformed Schwarzschild black hole. In more details, we have derived the frequency
  shift for two special cases: 1) When the photon is emitted
  from matter moving transverse to the observer ($\phi=0$
  or $\phi=\pi$). 2) When the matter is moving either directly
  towards or away from the observer ($\phi=\pm\frac{\pi}{2}$).
  We note that in contrast to the standard accretion into black hole,
  for regions smaller than $\tilde{r}_{min}=6$ there is a
  possibility of radiation so that the smallest frequency
  shifts are below the GR prediction. It seems defendable in the sense
  that by considering the rotation of black hole (Kerr metric)
  in calculations of the standard accretion disk, the photon frequency shift obtained from Schwarzschild
  black hole reduces by a small amount.

  \item As the third issue, we have studied the gravitational tidal forces around a GUP-deformed
  Schwarzschild black hole. Unlike the GR case, here depending on the sign of $\varepsilon$,
  we are dealing with some new physics. In the case $0<\varepsilon\leq1$, GUP-deformation
  term resists against tension or stretching produced via GR term in the radial direction
  $ \xi^{\hat{r}}$ as far as $r\leq\frac{3}{2}\mu$ where a transition from tension to
  compression situates. Also GUP-deformation term creates a repulsion against pressure or
  compression arising from GR term in the transverse directions ($ \xi^{\hat{\theta}}$
  and $ \xi^{\hat{\phi}}$) as far as $r\leq\mu$ where compression changes to tension. It
  is clear from the deformed line element (\ref{e2-1})
  that in the presence of a minimal length, the Schwarzschild radius depending on the
  $0<\varepsilon\leq1$ is located in the range $\frac{3}{2}\mu\leq r<2\mu$. Therefore,
  a freely falling particle passing the smallest of the Schwarzschild radius and entering
  the region $\mu<r<\frac{3}{2}\mu$, gets squeezed in all directions. As soon as its arrival to
  the region $0<r<\mu$, despite continuing the contraction in the radial direction, the particle
  expands in the transverse directions. In case of choosing the negative sign for the $\varepsilon$,
  the effect of tension and compression enhances in the radial and transverse directions, respectively.
  In this case as GR, there is no transition. By focusing on the aforementioned
  similarity between the deformed metric (\ref{e2-1}) and the Reissner-Nordstr\"{o}m metric with $\varepsilon\equiv\frac{e^2}{m^2}$, we can see
  that in the case of discarding the negative values of $\varepsilon$ the results derived here for the tidal forces
  are generally in agreement with those obtained within the RN metric, see \cite{LCB:2016}. The
  similarities between the metric (\ref{e2-1}) and Reissner-Nordstr\"{o}m metric as discussed here let us to say that:
  \emph{The dimensionless deformation parameter $\varepsilon$ in the line element
  (\ref{e2-1}) plays the same role that the charge to mass ratio $\frac{e}{m}$
  has within RN geometry.} 

  \item As the forth issue, we proposed the geodetic drift
  rate within the context of the GUP-deformed Schwarzschild
  geometry. Interestingly, our analysis reveals that in the case
  of choosing the positive values for the deformation parameter
  $0<\varepsilon\leq1$, GR prediction gets improved. In other words,
  deformation term in the line element (\ref{e2-1}) can lead to fill
  at least a part of the typical gap reported between GR prediction
  and data analysis released in Gravity Probe B experiment. As Table
  I shows, by setting $\varepsilon=0.1374$ one finds $\varphi_2=6.6020$
  arc-second/Yr in our setup which is much closer to the released value
  by \emph{Gravity Probe B} ($6.6018\pm0.0183$ arc-second/Yr) in comparison
  with \emph{GR+Kerr metric} result, $6.6061$ arc-second/Yr. As has been pointed out
  in Ref. \cite{Guo:2016}, the underlying GUP model implicitly is based on the assumption
  that the energy $E$ of the particle moving along the geodesic is equivalent to the
  Hawking temperature of the Schwarzschild black hole i.e. $E = T_h$. However,
  $T_h$ is much smaller than relevant energy of the Gravity Probe B. So, the question then arises: how this feature changes the relevant predictions for the GUP effects on the Gravity Probe B? The answer is that similar to what is reported in Ref. \cite{Fabio:2015},
  the modification derived for the underlying astrophysical phenomenon like the discussed issues in this paper,
  there is no explicit dependence on the energy of the particle moving along the geodesic. We note that in theories such as Gravity's Rainbow the situation is different.
\end{itemize}
\section{Acknowledgements}
We would like express our thanks to anonymous referee for his/her constructive comments on the original manuscript.


\begin{thebibliography}{10}
\bibitem{Hossenfelder:2013}
S. Hossenfelder, Living Rev. Relativity {\bf 16} (2013) 2.

\bibitem{Rovelli:1995}
C. Rovelli and L. Smolin, Nucl. Phys. B {\bf 442} (1995) 593.

\bibitem{Ashtekar:1997}
A. Ashtekar and J. Lewandowski, Class. Quantum Grav. {\bf 14} (1997) A55.

\bibitem{Gross:1988}
D. J. Gross, P. F. Mende, Nucl. Phys. B {\bf303} (1988) 407.

\bibitem{Amati:1989}
D. Amati, M. Ciafaloni and G. Veneziano, Phys. Lett. B {\bf 216} (1989) 41.

\bibitem{Amelino:2002}
 G. Amelino-Camelia, Int. J. Mod. Phys. D {\bf 11} (2002) 35.

\bibitem{Magueijo:2002}
J. Magueijo and L. Smolin, Phys. Rev. Lett. {\bf 88} (2002) 190403.

\bibitem{Seiberg:1999}
N. Seiberg and E. Witten, J. High Energy Phys. \textbf{9909} (1999) 032.

\bibitem{Connes:2000}
A. Connes, J. Math. Phys. \textbf{41} (2000) 3832.

\bibitem{Maggiore:1993}
M. Maggiore, Phys. Lett. B {\bf 304} (1993) 65.

\bibitem{Pikovski:2012}
I. Pikovski, M. R. Vanner, M. Aspelmeyer, M. Kim, C. Brukner, Nature Physics \textbf{8} (2012) 393.

\bibitem{Nozari:2011}
P. Pedram, K. Nozari and S. H. Taheri, JHEP \textbf{1103} (2011) 093.

\bibitem{Kempf:1995}
A. Kempf, G. Mangano and R. B. Mann, Phys. Rev. D {\bf 52} (1995) 1108.

\bibitem{Kempf:1997}
A. Kempf and G. Mangano,  Phys. Rev. D {\bf 55} (1997) 7909.

\bibitem{Nozari:2012}
K. Nozari and A. Etemadi, Phys. Rev. D {\bf85} (2012) 104029.

\bibitem{Hossain:2010}
G. M. Hossain, V. Husain and S. S. Seahra, Class. Quantum Grav. {\bf 27} (2010) 165013.

\bibitem{Acosta:2011}
 G. Chac\'{o}n-Acosta, L. Dagduga and H. A. Morales-T\'{e}cotla, AIP Conf. Proc. {\bf 1396} (2011) 99.

\bibitem{Gorji:2014}
M. A. Gorji, K. Nozari and B. Vakili, Phys. Lett. B {\bf 735} (2014) 62.

\bibitem{Nozari:2014}
M. A. Gorji, K. Nozari and B. Vakili, Phys. Rev. D {\bf 90} (2014) 044051.

\bibitem{Nozari:2015}
K. Nozari, M. Khodadi, M. A. Gorji, Europhys. Lett. {\bf 112} (2015) 60003.

\bibitem{Baru:1999}
F. Brau, J. Phys. A {\bf 32} (1999) 7691.

\bibitem{Das:2008}
 S. Das, E. Vagenas, Phys. Rev. Lett. {\bf 101} (2008) 221301.

\bibitem{Pedram:2011}
P. Pedram, K. Nozari, and S. H. Taheri, JHEP {\bf 1103} (2011) 093.

\bibitem{Ali:2011}
A. F. Ali, S. Das, E. Vagenas, Phys. Rev. D {\bf 84} (2011) 044013.

\bibitem{Cheng:2002}
L. N. Chang, D. Minic, N. Okamura and T. Takeuchi, Phys. Rev. D {\bf 66}(2002) 026003.

\bibitem{Nozari:2005}
K. Nozari, T. Azizi, Gen. Rel. Grav. {\bf 38}  (2006) 735.

\bibitem{Nozari:2008}
 K. Nozari, S. Akhshabi, Chaos Solitons Fractals {\bf 37} (2008) 324.

\bibitem{Pedram:2012}
P. Pedram, Phys. Lett. B {\bf 718} (2012) 638.

\bibitem{Fabio:2015}
F. Scardigli and R. Casadioc, Eur. Phys. J. C {\bf 75} (2015) 425.

\bibitem{Shapiro:2004}
Stuart L. Shapiro and Saul A. Teukolsky, \emph{``Black Holes, Whit Dwarfs, and Neutron Stars,
The Physics of Compact Objects"}, Wily-VCH Verlag GmbH and Co. KGaA, Veinheim (2004).

\bibitem{Bardeen:1972}
J. M. Bardeen, W. H. Press, S. A. Teukolsky, ApJ {\bf178} (1972) 347.

\bibitem{GP:2011}
C. W. F. Everitt at all, Phy. Rev. Lett {\bf 106} (2011) 221101.

\bibitem{Rajab:2010}
R. M. Gad, Astrophys. Space Sci. {\bf330} (2010) 107.
\bibitem{Ruffini:2011}
D. Pugliese, H. Quevedo, R. Ruffini, Phys. Rev. D {\bf83} (2011) 024021.
\bibitem{LCB:2016}
L. C. B. Crispino, A. Higuchi, L. A. Oliveira, E. S. de Oliveira,
Eur. Phys. J. C {\bf76} (2016)  168.
\bibitem{Guo:2016}
X. Guo, P. Wang, H. Yang, JCAP {\bf 062} (2016) 1605.

\end{thebibliography}
\end{document}